\documentclass[aps,prd,preprint,preprintnumbers,unsortedaddress,superscriptaddress,showpacs,nofootinbib,longbibliography]{revtex4-1}
\usepackage{subfigure}
\usepackage{graphicx}
\usepackage{epsfig}
\usepackage{amsmath}
\usepackage{amsfonts}
\usepackage{amssymb}
\usepackage{xcolor}
\usepackage{slashed}
\usepackage{lipsum}
\usepackage{epstopdf}
\usepackage{hyperref}
\usepackage{float}
\usepackage{wrapfig}
\usepackage[utf8]{inputenc}
\hypersetup{colorlinks = true,
           allcolors = blue}

\newcommand{\tento}[1]{\times 10^{#1}}

\newcommand{\mL}{\mathcal{L}}
\newcommand{\mM}{\mathcal{M}}

\newcommand{\mev}{{\rm MeV}}
\newcommand{\gev}{{\rm GeV}}

\newcommand{\itp}{\affiliation{CAS Key Laboratory of Theoretical Physics, Institute of Theoretical Physics, Chinese Academy of Sciences,  Zhong Guan Cun East Street 55, Beijing 100190, China}}

\allowdisplaybreaks[1]
\usepackage{version}	
\begin{document}
\title{Role of the triangle mechanism in the $\Lambda_b\rightarrow \Lambda_c\pi^-f_0(980)$ reaction}
\date{\today}
\author{Shuntaro Sakai}
\email{shsakai@mail.itp.ac.cn}
\itp

\begin{abstract} 
 We investigate the $\Lambda_b\to\Lambda_c\pi^-f_0(980)$ production with a $f_0(980)$ decay into $\pi^+\pi^-$ via the $K^{*0}K^-K^+$ and $K^{*-}K^0\bar{K}^0$ triangle loops.
 These loops produce a peak around 1.42~GeV in the $\pi^-f_0(980)$ invariant mass distribution,
 which is the same mechanism as the one considered to explain the $a_1(1420)$ peak.
 In the $\pi^+\pi^-$ distribution obtained by fixing the $\pi^-f_0(980)$ invariant mass to some values, a clear peak of $f_0(980)$ is seen,
 and the $\pi^-f_0(980)$ distribution has a peak around $M_{\pi^-f_0}=1.42$~GeV, which is caused by the triangle mechanism of the $K^*\bar KK$ loop.
 The branching ratio of $\Lambda_b\rightarrow\Lambda_c\pi^-f_0(980)$ with $f_0(980)\rightarrow\pi^+\pi^-$ by the triangle mechanism, 
 obtained by integrating the $\pi^-f_0(980)$ distribution from 1 to 1.6~GeV, is estimated to be the order $10^{-4}$.
 Future measurements of the $\Lambda_b\to\Lambda_c\pi^-f_0(980)$ branching ratio and the $\pi^-f_0(980)$ invariant mass distribution predicted in this work would give further clues to clarify the nature of the $a_1(1420)$ peak.
\end{abstract}
\maketitle

\section{Introduction\label{sec_introduction}}
The role of triangle singularities (TSs) in hadronic reactions has been investigated for a long time.
A general discussion on the emergence of singularities from loop amplitudes was given by Landau~\cite{Landau:1959fi},
and a physical picture of the singularity from triangle loops was provided in Ref.~\cite{Coleman:1965xm} and is known as the Coleman-Norton theorem;
the TSs can show up when all the internal particles are on shell, the momenta of the particles in the loop are collinear, and the process can occur at the classical level.
One can find a refined formulation and an intuitive picture of the TS in Ref.~\cite{Bayar:2016ftu} (see also Ref.~\cite{Guo:2019twa} for a recent review of the TS).

\begin{figure}[t]
 \centering
 \includegraphics[width=5cm]{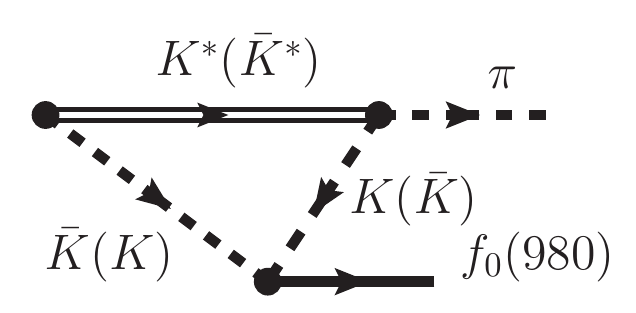}
 \caption{Triangle diagram relevant to the $\pi f_0(980)$ production.}
 \label{fig:tl}
\end{figure}

One interesting manifestation of the TS is the $\eta(1405/1475)\rightarrow\pi^0f_0(980)$ decay.
An anomalously large production of the $\pi^0f_0(980)$ decay mode of $\eta(1405/1475)$, which is forbidden by isospin symmetry, was reported by the BESIII~Collaboration~\cite{BESIII:2012aa},
and that large amount production rate and the narrow $f_0(980)$ line shape in the $\pi\pi$ distribution, 
which is the order of the mass difference of the charged and neutral kaons due to the isospin symmetry breaking, 
are explained well by the triangle mechanism~\cite{Wu:2011yx,Aceti:2012dj,Wu:2012pg,Achasov:2015uua} (see also Ref.~\cite{Achasov:2019vcs} for a review article).
The triangle diagram considered in the work is composed of $K^*\bar KK$ and its charge conjugation shown in Fig.~\ref{fig:tl}.
The triangle loop diagram of Fig.~\ref{fig:tl} has a singularity around $1.42~\gev$, which is in the $\eta(1405/1475)$ mass region.
In the process, the $K^*\bar K$ pair is produced by $\eta(1405/1475)$ first, the $K^*$ decays into $\pi K$, and the $K\bar K$ couple to $f_0(980)$.
The position of the singularity can be obtained with a formula given in Ref.~\cite{Bayar:2016ftu}.
The triangle singularity plays an essential role in this process
because the position of the TS is sensitive to the masses of the particles and the mass difference of $K$ and $K^*$ involved in the triangle loop introduces the isospin violation in this process.
In practice, the singularity is turned into a peak by the width of the internal particles, and the detailed study on the width effect was done in Ref.~\cite{Du:2019idk}.
In Refs.~\cite{Sakai:2017iqs,Liang:2017ijf}, some other processes were studied for further investigation of the anomalous enhancement of the isospin-forbidden $\pi^0f_0(980)$ production by the triangle mechanism.
The TS of the $K^*\bar{K}K$ loop was mentioned in Ref.~\cite{Schmid:1967ojm}, and
the possible role of the $K^*\bar{K}K$ triangle loop has been investigated in many processes~\cite{Wu:2011yx,Aceti:2012dj,Wu:2012pg,Achasov:2015uua,Achasov:1989ma,Aceti:2015zva,Ketzer:2015tqa,Du:2019idk,Sakai:2017iqs,Liang:2017ijf,Aceti:2016yeb,Debastiani:2016xgg,Pavao:2017kcr,Dai:2018rra,Jing:2019cbw}.

One important aim to study the TS is to clarify the origin of peaks in the invariant mass distribution.
The peak of the TS has purely kinematical origin and cannot be associated with a resonant state.
A peak of $a_1(1420)$, which is in the $p$-wave $\pi^-f_0(980)$ mode in the $\pi^-p\to \pi^+\pi^-\pi^-p$ reaction,
was found by the COMPASS Collaboration~\cite{Adolph:2015pws,Akhunzyanov:2018lqa} followed by the studies on the properties of $a_1(1420)$~\cite{Chen:2015fwa,Gutsche:2017oro,Gutsche:2017twh,Sundu:2017xct,Murakami:2018spb}
and the work on the explanation of the peak focusing on its production mechanism~\cite{Basdevant:2015wma} 
(see Ref.~\cite{Ketzer:2019wmd} for a recent review article, and see also a mini review for mesons in the $1400~\mev$ region in the Particle Data Group (PDG)~\cite{Tanabashi:2018oca}).
A possible understanding of the peak with the triangle mechanism was suggested in Refs.~\cite{Ketzer:2015tqa,Aceti:2016yeb}.
The position of the singularity around $1.42~\gev$ stemming from the $K^*\bar{K}K$ loop coincides with the peak position of $a_1(1420)$.
Despite the attempts to clarify the nature of the peak, a significant difference of the resonance and TS scenarios of the $a_1(1420)$ peak has not been found in the partial wave analysis so far~\cite{Ketzer:2019wmd}.
Some predictions based on the TS scenario of the $a_1(1420)$ peak were made in Refs.~\cite{Pavao:2017kcr,Dai:2018rra} in the $B$ and $\tau$ decays.

\begin{figure}[t]
 \centering
 \includegraphics[width=12cm]{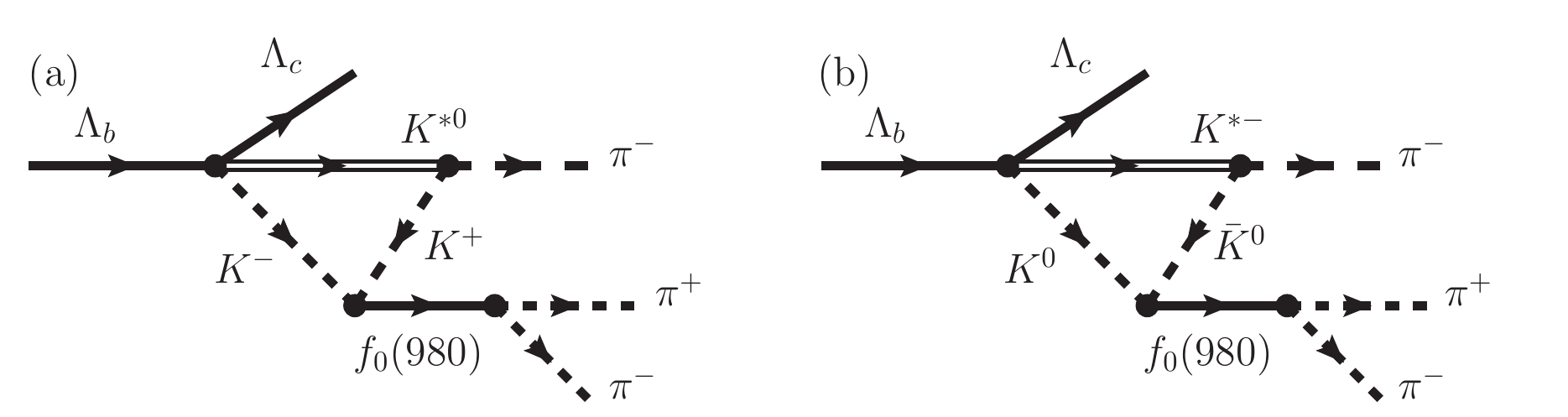}
 \caption{Triangle loops for the $\Lambda_b\rightarrow\Lambda_c\pi^-f_0(980)$ process.}
 \label{fig-diagram}
\end{figure}

In this work, we study the $\Lambda_b\rightarrow\Lambda_c\pi^-f_0(980)$ with $f_0(980)\rightarrow\pi^+\pi^-$
via the triangle mechanism of the $K^*\bar{K}K$ loop producing a peak around 1.42~GeV in the $\pi^-f_0(980)$ distribution.
We show in Fig.~\ref{fig-diagram} the diagram of the $K^*\bar KK$ triangle loop contributing to the $\Lambda_b\to\Lambda_c\pi^-f_0(980)$ process.
Apart from the $K^*\bar K$ production part, the mechanism producing a peak around 1.42~GeV is identical to the one considered in Refs.~\cite{Ketzer:2015tqa,Aceti:2016yeb} for the $a_1(1420)$ peak.
With known theoretical and experimental information, we make the predictions on the branching ratio and the invariant mass distribution in the $\Lambda_b\to \Lambda_c\pi^-f_0(980)$ reaction.
Such predictions, including the triangle mechanism, will be important for future experiments to clarify the nature of $a_1(1420)$ as an unavoidable peak of the kinematical effect.

\section{Setup\label{sec_setup}}
The diagrams we consider in this study are shown in Fig.~\ref{fig-diagram}.
Here, we focus on the diagram Fig.~\ref{fig-diagram}(a) in which a loop is formed by $K^{*0}K^-{K}^+$;
$\Lambda_b$ first decays into $\Lambda_cK^{*0}K^-$, and subsequently $K^{*0}$ turns into the $\pi^-{K}^+$ with a merging of $K^+$ and ${K}^-$ to give $f_0(980)$.
The $f_0(980)$ finally decays into a $\pi^+\pi^-$ pair.
Strictly speaking, this $K^*\bar KK$ loop for the $\pi^-f_0(980)$ production does not have the TS because the mass of $f_0(980)$ is slightly below the $K\bar K$ threshold;
i.e., the $K\bar K$ in the loop cannot be on shell.
However, in the distribution the remnant of the TS would be still expected due to the width of the particles.
For example, by putting the $f_0(980)$ mass slightly above the $K\bar K$ threshold, with the formula in Ref.~\cite{Bayar:2016ftu}, the loop amplitude produces a TS around $M_{\pi^-f_0}=1.42~\gev$ in the $\pi^-f_0(980)$ invariant mass distribution.

In this section, the amplitude needed to evaluate the diagram in Fig.~\ref{fig-diagram},
the $\Lambda_b\to\Lambda_cK^*\bar K$, $K^*\to\pi K$, and $K\bar K\to \pi^+\pi^-$ transition amplitudes, will be explained,
and the amplitude of the $\Lambda_b\to \Lambda_c\pi^-f_0(980)$ decay with $f_0(980)\to\pi^+\pi^-$ will be given at the end of this section.

\subsection{$\Lambda_b\to\Lambda_cK^{*}\bar K$ amplitude}
\begin{figure}[t]
 \centering
 \includegraphics[width=12cm]{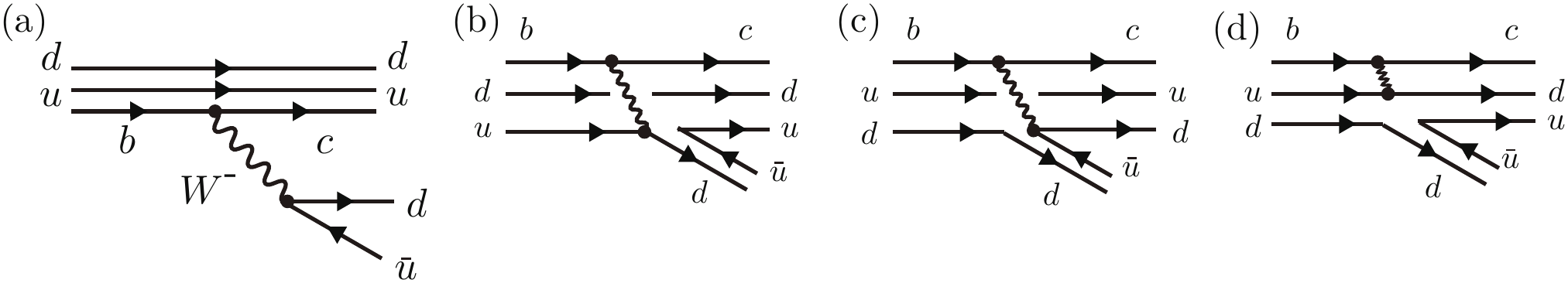}
 \caption{Quark-level diagrams for the $\Lambda_b$ decay.}
 \label{fig_weak-diagram}
\end{figure}

First, we consider the $\Lambda_b\rightarrow \Lambda_c K^{*0}{K}^-$ amplitude.
At present, the data of the $\Lambda_b\rightarrow \Lambda_c K^{*0}{K}^-$ decay, such as the branching fraction or the Dalitz plot distribution, are not available;
then, we make a microscopic derivation of the $\Lambda_b\to\Lambda_cK^{*}\bar K$ amplitude with some approximations.
Some possible diagrams for the $\Lambda_b\to \Lambda_cK^*\bar K$ at quark level are depicted in Fig.~\ref{fig_weak-diagram}.
In this calculation, we take account of the diagram Fig.~\ref{fig_weak-diagram}(a) with the external $W^-$ emission, which is favored in terms of the color counting~\cite{Chau:1982da}, 
and the diagrams with different topology shown in diagrams (b), (c), and (d) in the figure, suppressed in terms of the color counting, are not considered,
and that can give, in general, the uncertainties of the order of a few tens of percents.\footnote{Changing the $u$ quark with the $d$ quark, the diagrams (b), (c), and (d) in Fig.~\ref{fig_weak-diagram} can lead to the $\Sigma_c$ production, which is reported to be small~\cite{Tanabashi:2018oca}.
In addition, when one sees the $B\to KD^*\bar D^*$ branching fraction in the PDG~\cite{Tanabashi:2018oca}, 
the branching fraction of the color favored process is about ten times larger than that of the color disfavored one~\cite{Sakai:2020ucu}.
These facts would imply the small corrections from the color disfavored processes.}

For the baryonic part of the $\Lambda_b\to \Lambda_c W^-$ transition, $B^\mu$,
\begin{align}
  B^\mu=&\left<\Lambda_c\left|\bar{c}\gamma^\mu(1-\gamma_5)b\right|\Lambda_b\right>,\label{eq-Bmu} 
\end{align}
we use a result of the QCD sum rule~\cite{Azizi:2018axf}.
The $\Lambda_b\rightarrow\Lambda_c$ transition amplitude is parametrized as
\begin{align}
 \left<\Lambda_c\left|V^\mu\right|\Lambda_b\right>=\left<\Lambda_c\left|\bar{c}\gamma^\mu b\right|\Lambda_b\right>
 =&\bar{u}_{\Lambda_c}\left[F_1(q^2)\gamma^\mu+F_2(q^2)v^\mu+F_3(q^2)v'^\mu\right]u_{\Lambda_b},\\
 \left<\Lambda_c\left|A^\mu\right|\Lambda_b\right>=\left<\Lambda_c\left|\bar{c}\gamma^\mu\gamma_5 b\right|\Lambda_b\right>
 =&\bar{u}_{\Lambda_c}\left[G_1(q^2)\gamma^\mu+G_2(q^2)v^\mu+G_3(q^2)v'^\mu\right]\gamma_5u_{\Lambda_b},
\end{align}
where $v^\mu(v'^\mu)=p^\mu_{\Lambda_b(\Lambda_c)}/m_{\Lambda_b(\Lambda_c)}$ is the 4-velocity of $\Lambda_b$ $(\Lambda_c)$, $q^\mu=p^\mu_{\Lambda_b}-p^\mu_{\Lambda_c}$, 
$\mathcal{F}(q^2)=F_i(q^2)$ or $G_i(q^2)$ $(i=1,2,3)$ is parametrized as
\begin{align}
 \mathcal{F}(q^2)=\frac{\mathcal{F}(0)}{1-\xi_1\frac{q^2}{m_{\Lambda_b}^2}+\xi_2\frac{q^4}{m_{\Lambda_b}^4}+\xi_3\frac{q^6}{m_{\Lambda_b}^6}+\xi_4\frac{q^8}{m_{\Lambda_b}^8}},\label{eq:ffdef}
\end{align}
with $\mathcal{F}(0)$ and $\xi_i$ given in Ref.~\cite{Azizi:2018axf}.
For later purposes, we give the spin sum and average of the baryonic part $\bar{B}^{\mu\nu}=\overline{\sum}\sum B^\mu B^{*\nu}$ (quantities with overline denote the spin summed and averaged ones in the following):
\begin{align}
 \bar{B}^{\mu\nu}=&\frac{1}{2}\left[g^{\mu\nu}\left\{(F_1^2-G_1^2)-(F_1^2+G_1^2)w\right\}+(F_1^2+F_1F_2)(v^\mu v'^\nu+v'^\mu v^\nu)\right.\notag\\
 &+2F_1F_2v^\mu v^\nu +F_1F_3(v^\mu v'^\nu+v'^\mu v^\nu+2v'^\mu v'^\nu)\notag\\
 &+w\left\{(F_2^2+G_2^2)v^\mu v^\nu +(F_2F_3+G_2G_3)(v^\mu v'^\nu+v'^\mu v^\nu)+(F_3^2+G_3^2)v'^\mu v'^\nu\right\}\notag\\
 &+F_2^2v^\mu v^\mu +F_2F_3(v^\mu v'^\nu+v'^\mu v^\nu)+F_3^2v'^\mu v'^\nu\notag\\
 &+G_1^2(v^\mu v'^\nu+v'^\mu v^\nu)-G_1G_2(v^\mu v'^\nu+v'^\mu v^\nu)+2G_1G_2v^\mu v^\nu \notag\\
 &-2G_1G_3 v'^\mu v'^\nu +G_1G_3(v^\mu v'^\nu+v'^\mu v^\nu)-G_2^2v^\mu v^\nu\notag\\
 &\left. -G_2G_3(v^\mu v'^\nu+v'^\mu v^\nu)-G_3^2v'^\mu v'^\nu+2iF_1G_1\epsilon^{\mu\nu\rho\sigma}v_\rho v'_\sigma\right]\label{eq-spin-sum-av}
\end{align}
with $w=v\cdot v'$.

Let us move to the mesonic part producing $K^{*0}K^-$ or $K^{*-}K^0$, which is denoted by $J_\mu$.
Here, we assume the $a_1^-(1260)$ dominance for the ${K}^{*0}K^{-}$ and $K^{*-}{K}^0$ production and the effects of the rescatterings of the other pairs in the final-state particles are ignored.
The observation of $\Lambda_b\to\Lambda_ca_1^-(1260)$ is reported in Ref.~\cite{Abreu:1996mi}, and a peak around $1.3~\gev$ which may be associated with $a_1(1260)$ is seen in the $\pi^-\pi^+\pi^-$ distribution of $\Lambda_b\rightarrow\Lambda_c\pi^-\pi^+\pi^-$~\cite{Aaij:2011rj},
and we expect a large portion of $\Lambda_b\to\Lambda_c a_1^-(1260)$ in the $K^*\bar K$ production 
by taking into account a fairly strong coupling of $a_1(1260)$ to $K^*\bar{K}$ obtained theoretically in Refs.~\cite{Roca:2005nm,Zhou:2014ila}.
Then, $J_\mu$ represents the amplitude of the $W^-\to a_1^-(1260)\to K^{*0}K^-$ or $K^{*-}K^0$ transition here.
The effect of $a_1(1260)$ on the $\pi^-f_0(980)$ distribution will be checked later.

We write the conversion amplitude from $W^-$ to $a_1^-(1260)$ as
\begin{align}
 -it_{W^-,a_1^-}=&\frac{ig_WV_{ud}}{2}f_{a_1}m_{a_1}\epsilon_{W^-}\cdot\epsilon_{a_1^-}^*\label{eq:wa1}
\end{align}
with $V_{ud}$ and $g_W$ being the element of the Cabibbo-Kobayashi-Maskawa quark-mixing matrix and the coupling constant of the weak interaction, respectively,
and the 
$a_1^-(1260)\rightarrow K^{*0}{K}^-$
amplitude is written as
\begin{align}
 \begin{split}
  -it_{a_1^-,{K}^{*0}{K}^-}=&g_1\epsilon_{a_1^-}\cdot\epsilon_{{K}^{*0}}^*,
 \end{split}\label{eq-v1}
\end{align}
where we take into account the amplitude with the lowest angular momentum which gives the dominant contribution in low energies.
In the case of the $K^{*-}K^0$ pair in the final state, we just need an additional minus sign.

With the $\Lambda_b\to\Lambda_c$ form factor $B^\mu$ in Eq.~\eqref{eq-Bmu}, 
the $\Lambda_b\to \Lambda_cK^{*0}K^-$ transition amplitude given by the external $W^-$ emission process 
is written as
\begin{align}
 -i\mathcal{M}_{\Lambda_b,\Lambda_cK^{*0}{K}^-}=&\left(\frac{ig_W}{2\sqrt{2}}V_{cb}\right)B^\mu\frac{i\left(-g_{\mu\nu}+\frac{q_\mu q_\nu}{m_W^2}\right)}{q^2-m_W^2+i\epsilon}\left(\frac{ig_WV_{ud}}{2}f_{a_1}m_{a_1}\right)\notag\\
 &\cdot\frac{i\left(-g^{\nu\rho}+\frac{q^\nu q^\rho}{m_{a_1}^2}\right)}{q^2-m_{a_1}^2+im_{a_1}\Gamma_{a_1}}(+g_1)(\epsilon_{K^{*0}}^{*})_\rho\notag\\
 \sim&g_1G_FV_{cb}V_{ud}B^\mu g_{\mu\nu}\frac{f_{a_1}m_{a_1}\left(-g^{\nu\rho}+\frac{q^\nu q^\rho}{m_W^2}\right)}{q^2-m_{a_1}^2+im_{a_1}\Gamma}(\epsilon_{K^{*0}}^{*})_\rho\notag\\
 \equiv&G_FV_{ud}V_{cb}B^\mu J_\mu,\label{eq:lbtolckstkb}\\
 J^\mu=&g_1G_{a_1}^{\mu\nu}(\epsilon_{K^{*0}}^{*})_\nu,~
 G_{a_1}^{\mu\nu}=\frac{f_{a_1}m_{a_1}\left(-g^{\mu\nu}+\frac{q^\mu q^\nu}{m_{a_1}^2}\right)}{q^2-m_{a_1}^2+im_{a_1}\Gamma_{a_1}},
\end{align}
by taking the leading-order term of $1/m_W^2$.
$G_F=g_W^2/(4\sqrt{2}m_W^2)$ is the Fermi coupling constant.
The $\Lambda_b\to \Lambda_cK^{*-}K^0$ amplitude has the opposite sign relative to Eq.~\eqref{eq:lbtolckstkb}, which comes from the different sign of the $a_1^-(1260)\to K^{*0}K^-$ and $K^{*-}K^0$ vertices.

We fix the parameters in Eqs.~\eqref{eq:wa1} and \eqref{eq-v1} with the $\tau^-\rightarrow\nu_\tau K^{*0}K^-$ partial width given in the PDG~\cite{Tanabashi:2018oca},
assuming the production is dominated by the $a_1(1260)$ meson.
In the spectral function of $\tau\to 3\pi\nu_\tau$~\cite{Schael:2005am,Davier:2013sfa}, one can see a significant peak at $1.2~\gev^2$.
The $K^-K^+\pi^-$ distribution of the $\tau^-\to \nu_\tau K^{-}K^+\pi^-$ decay, in which a large amount of $K^+\pi^-$ comes from $K^{*0}$, is available~\cite{Barate:1997ma}, but the data are not enough for the analysis.
The matrix element of the $\tau^-\to\nu_\tau K^{*0}K^-$ decay is written as
\begin{align}
 -i\mM_\tau=&\left(\frac{ig_W}{2\sqrt{2}}\right)\bar{u}_{\nu}\gamma^\mu(1-\gamma_5)u_\tau\frac{i\left(-g_{\mu\nu}+\frac{q_\mu q_\nu}{m_W^2}\right)}{q^2-m_W^2+i\epsilon}\left(\frac{ig_WV_{ud}}{2}f_{a_1}m_{a_1}\right)\notag\\
 &\cdot\frac{i\left(-g^{\nu\rho}+\frac{q^\nu q^\rho}{m_{a_1}^2}\right)}{q^2-m_{a_1}^2+im_{a_1}\Gamma_{a_1}}
\left(+g_1\right)(\epsilon^*_{{K}^{*0}})_\rho\notag\\
 \sim&g_1G_FV_{ud}\bar{u}_{\nu}\gamma^\mu(1-\gamma_5)u_\tau g_{\mu\nu}
\frac{f_{a_1}m_{a_1}\left(-g^{\nu\rho}+\frac{q^\nu q^\rho}{m_{a_1}^2}\right)}{q^2-m_{a_1}^2+im_{a_1}\Gamma_{a_1}}
(\epsilon^*_{{K}^{*0}})_\rho\\
 \equiv&G_FV_{ud}\mL^\mu J_\mu,\label{eq:tauamp}\\
 \mL^\mu=&\bar{u}_{\nu}\gamma^\mu(1-\gamma_5)u_\tau.
\end{align}
For the $a_1(1260)$ mass and width, the central value of the PDG~\cite{Tanabashi:2018oca} is used; $m_{a_1}=1.23~\gev$ and $\Gamma_{a_1}=0.425~\gev$.\footnote{The peak in the spectral function in Refs.~\cite{Schael:2005am,Davier:2013sfa} is a bit lower than the $a_1(1260)$ mass from the PDG~\cite{Tanabashi:2018oca}.
See, e.g., Ref.~\cite{Mikhasenko:2018bzm} for a recent study on the $a_1(1260)$ meson in the $\tau$ decay,
and see also Refs.~\cite{Urban:2001ru,GomezDumm:2003ku,Lutz:2003fm,Roca:2005nm,Wagner:2008gz,Nagahiro:2011jn,Kamano:2011ih,Parganlija:2012fy,Roca:2012rx,Lang:2014tia,Zhou:2014ila,Zhang:2018tko,Osipov:2018iah,Murakami:2018spb,Sadasivan:2020syi} and the references therein for the works concerning the $a_1(1260)$ properties.}
The spin sum and average of the leptonic part $\mL^\mu$ is given by
\begin{align}
 \mL^{\alpha\beta}\equiv\overline{\mL^\alpha \mL^{*\beta}}=&\frac{4}{4m_\tau m_{\nu}}[p_\tau^\alpha p_\nu^\beta+p_\tau^\beta p_\nu^\alpha-g^{\alpha\beta}(p_\tau\cdot p_\nu)+i\epsilon^{\alpha\beta\rho\sigma}(p_{\tau})_\rho (p_{\nu})_\sigma],
\end{align}
with $p_\tau$ and $p_\nu$ being the momenta of the $\tau$ lepton and neutrino, respectively.
Then, the matrix element squared with the spin sum and average is
\begin{align}
 \overline{|\mM_\tau|^2}=&{g_1^2G_F^2V_{ud}^2}
 \mL_{\mu\nu}G_{a_1}^{\mu\mu'}G_{a_1}^{*\nu\nu'}\left[-g_{\mu'\nu'}+\frac{(p_{K^{*0}})_{\mu'}(p_{K^{*0}})_{\nu'}}{m_{K^{*0}}^2}\right],\label{eq:tmp1}
\end{align}
and the differential width is 
\begin{align}
 \frac{d^2\Gamma_{\tau^-\rightarrow\nu_\tau K^{*0}K^-}}{dM_{K^-\nu_\tau}^2dM_{K^{*0}K^-}^2}=&\frac{4m_\tau m_\nu}{(2\pi)^3 32m_\tau^3}\overline{|\mM_\tau|^2}.
\end{align}
In practice, we do not need to fix $f_{a_1}$ and $g_1$ independently because a product $f_{a_1}m_{a_1}g_1$ appears in the amplitudes Eqs.~\eqref{eq:lbtolckstkb} and \eqref{eq:tauamp}.
Finally, with the partial width of $\tau^-\to \nu_\tau K^{*0}K^-$, the product $f_{a_1}m_{a_1}g_1$ is fixed to $f_{a_1}m_{a_1}g_1=1.0~\gev^3$.

\subsection{$K^*\to \pi K$ and $\bar K^*\to \pi \bar K$ amplitudes}
We move to the $K^{*-}\rightarrow\pi^-\bar{K}^0$ and $K^{*0}\rightarrow\pi^-K^+$ amplitudes.
The $p$-wave amplitude of a vector meson decaying into two pseudoscalar mesons can be obtained from the effective Lagrangian \cite{Bando:1984ej,Bando:1987br,Meissner:1987ge,Nagahiro:2008cv,Jing:2019cbw},
\begin{align}
 \mL_{VPP}=-{i\tilde{g}}\left<V^\mu[ P,\partial_\mu P]\right>,
\end{align}
with
\begin{align}
 P = 
 \begin{pmatrix} 
  \frac{\pi^0}{\sqrt{2}}+\frac{1}{\sqrt{6}}\eta&\pi^+&K^+ \\ 
  \pi^-&-\frac{\pi^0}{\sqrt{2}}+\frac{1}{\sqrt{6}}\eta&K^0\\ 
  K^-&\bar{K}^0&-\sqrt{\frac{2}{3}}\eta
 \end{pmatrix},~
 V_\mu = 
 \begin{pmatrix}
  \frac{\rho^0+\omega}{\sqrt{2}}&\rho^+&K^{*+} \\ 
  \rho^-&\frac{-\rho^0+\omega}{\sqrt{2}}&K^{*0}\\ 
  K^{*-}&\bar{K}^{*0}&\phi
 \end{pmatrix}_\mu.
\end{align}
The brackets $\left<...\right>$ stand for the trace of the flavor SU(3) matrices.
From this Lagrangian, the amplitudes of $K^{*0}\to\pi^-K^+$ and $K^{*-}\to\pi^-\bar{K}^0$ are given by
\begin{align}
 \begin{split}
  -it_{K^{*0},\pi^-K^+}=&+{i\tilde{g}}{\epsilon}_{K^{*0}}\cdot\left({p}_{\pi^-}-{p}_{K^+}\right),\\
  -it_{K^{*-},\pi^-\bar{K}^0}=&-{i\tilde{g}}{\epsilon}_{K^{*-}}\cdot\left({p}_{\pi^-}-{p}_{\bar{K}^0}\right).
 \end{split}
 \label{eqkst}
\end{align}
We fix the parameter $\tilde{g}$ for the coupling of $K^*\rightarrow\pi K$ with the isospin averaged mass and width of mesons;
\begin{align}
 \Gamma_{K^{*}}=&\frac{2\tilde{g}^2p_{K}^3}{8\pi m_{K^{*}}^2},~
 p_K=\frac{1}{2m_{K^*}}\lambda^{1/2}(m_{K^*}^2,m_K^2,m_\pi^2),\label{eq:Kstwidth}
\end{align}
with the K\"all\`an function $\lambda(x,y,z)=x^2+y^2+z^2-2xy-2yz-2zx$,
and Eq.~\eqref{eq:Kstwidth} leads to  $\tilde{g}=4.5$.

\subsection{$K\bar K\to \pi^+\pi^-$ scattering amplitude}
For the scattering $t$ matrix of the $K\bar K$ to a meson pair $MM'$, $t_{MM',K\bar K}$,
we use the amplitude calculated in the framework of chiral unitary approach.
The $\pi\pi$-$K\bar{K}$-$\pi\eta$ coupled-channel system around $1~\gev$ was studied in this framework in Ref.~\cite{Oller:1997ti}
with a particular interest in the $f_0$ and $a_0$ resonances followed by the studies with similar approaches~\cite{Kaiser:1998fi,Locher:1997gr,Oller:1998hw}\footnote{The scalar mesons around 1~GeV have been studied for a long time, and many studies were devoted for it from various viewpoints, e.g., as done in Refs.~\cite{Jaffe:1976ig,Weinstein:1990gu,Janssen:1994wn,Baru:2003qq} (see also Ref.~\cite{Guo:2017jvc} for a review article).} and the applications to many reactions.
In this work, we follow the setup of Ref.~\cite{Liang:2014tia};
the $f_0(980)$ resonance is dynamically generated as a result of the nonperturbative meson-meson interaction, and it is found in Refs.~\cite{Liang:2014tia,Liang:2016hmr,Dias:2016gou} that the line shape of the $\pi^+\pi^-$ invariant mass distribution around $1~\gev$ is described fairly well.
The amplitude is given by the scattering equation
\begin{align}
 t_{i,j}=[(1-vg)^{-1}v]_{i,j}
\end{align}
with $i,j=\pi^+\pi^-,\pi^0\pi^0,K^+K^-,K^0\bar{K}^0,\eta\eta$.
The interaction kernel $v$ comes from the $s$-wave part of the leading-order chiral Lagrangian,
and $g$ is the meson-meson loop function with cutoff regularization given in Ref.~\cite{Liang:2014tia},
where the cutoff parameter $\Lambda$ for $g$ is chosen to be $0.6~\gev$.

\subsection{$\Lambda_b\to \Lambda_c\pi^-f_0(980)$ amplitude via the $K^*\bar KK$ loop}
\begin{figure}[t]
 \centering
 \includegraphics[width=6cm]{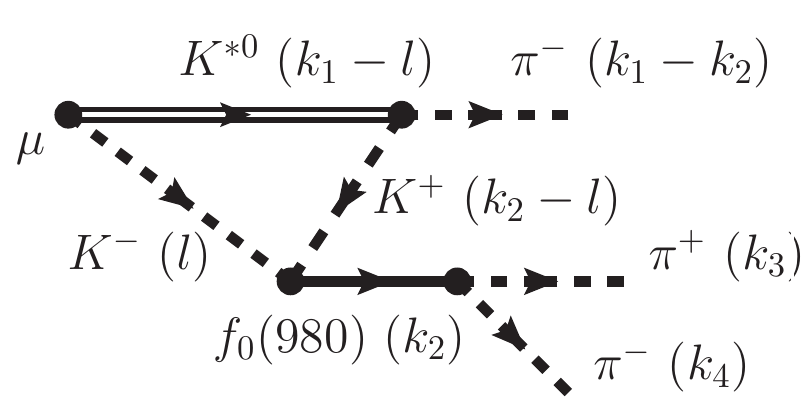}
 \caption{The $K^{*0}K^-{K}^+$ loop diagram for the $\pi^-f_0(980)$ transition amplitude $T_\mu$.}
 \label{fig2}
\end{figure}

Combining the amplitudes given above, we can obtain the loop amplitude given by the diagram in Fig.~\ref{fig2}, which is denoted by $T_\mu$.
We can write the amplitude given by the diagram in Fig.~\ref{fig2} with the $K^{*0}K^-{K}^+$ triangle loop (a meson pair $MM'$ in the final state comes from the $K^+{K}^-$ rescattering) as follows:
 \begin{align}
 T_\mu=&
 -i\tilde{g}t_{MM',K^+{K}^-}\int\frac{d^4l}{(2\pi)^4}\frac{\left(-g_{\mu\nu}+\frac{(k_1-l)_\mu (k_1-l)_\nu}{m_{K^{*0}}^2}\right)(k_1-2k_2+l)^\nu}{[l^2-m_{K^-}^2+i\epsilon][(k_1-l)^2-m_{K^{*0}}^2+i\epsilon][(k_2-l)^2-m_{{K}^+}^2+i\epsilon]}\notag\\
\equiv&-i\tilde{g}L_\mu(K^{*0}K^-{K}^+)t_{MM',K^+{K}^-},
 \end{align}
with
\begin{align}
 L_\mu(K^{*0}K^-{K}^+)
 =&\int\frac{d^4l}{(2\pi)^4}\frac{\left(-g_{\mu\nu}+\frac{(k_1-l)_\mu(k_1-l)_\nu}{m_{K^{*0}}^2}\right)(k_1-2k_2+l)^\nu}{[l^2-m_{K^-}^2+i\epsilon][(k_1-l)^2-m_{K^{*0}}^2+i\epsilon][(k_2-l)^2-m_{{K}^+}^2+i\epsilon]}\notag\\
 =&\int\frac{d^4l}{(2\pi)^4}\frac{1}{[l^2-m_{K^-}^2+i\epsilon][(l+k_1)^2-m_{K^{*0}}^2+i\epsilon][(l+k_2)^2-m_{{K}^+}^2+i\epsilon]}\notag\\
 &\cdot\left[-(k_1-2k_2)_\mu+\frac{k_1\cdot(k_1-2k_2)}{m_{K^{*0}}^2}k_{1\mu}+\left(1+\frac{k_1\cdot(k_1-2k_2)}{m_{K^{*0}}^2}\right)l_\mu\right.\notag\\
 &\left.-\frac{2k_2\cdot l}{m_{K^{*0}}^2}k_{1\mu}-\frac{2k_2\cdot l}{m_{K^{*0}}^2}l_\mu-\frac{l^2}{m_{K^{*0}}^2}k_{1\mu}-\frac{l^2}{m_{K^{*0}}^2}l_\mu\right].\label{eq:defLmu}
\end{align}
A library for the one-loop integrals, LoopTools, is used~\cite{Hahn:1998yk}.
In Eq.~\eqref{eq:defLmu}, 
the momenta $k_1$ and $k_2$ are defined as in Fig.~\ref{fig2}.
Now, the renormalization scale, $\mu$, associated with the divergence of the loop integral is fixed to be $1~\gev$, and the change of this parameter to $\mu=0.5$ or $1.5~\gev$ gives just a tiny difference.
The width effect of the $K^{*0}$ meson in the loop is included by replacing the squared mass of the $K^{*0}$, $m_{K^{*0}}^2$, with $m_{K^{*0}}^2-im_{K^{*0}}\Gamma_{K^{*0}}$ in this study.

Then, with the part of the $\Lambda_b\to \Lambda_cK^{*0}K^-$ transition given in Eq.~\eqref{eq:lbtolckstkb}, the $\Lambda_b\to\Lambda_c\pi^-f_0(980);f_0(980)\to MM'$ amplitude via the $K^{*0}K^-K^+$ triangle loop is written as
\begin{align}
 -i\mM_{\Lambda_b,\Lambda_c\pi^-MM'}^{(K^{*0}K^-K^+)}=&-i\tilde{g}g_1G_FV_{ud}V_{cb}B_\mu G_{a_1}^{\mu\nu}t_{MM',K^+K^-}L_\nu(K^{*0}K^-K^+).
\end{align}
The amplitude of the $K^{*-}K^0\bar K^0$ loop is obtained by just changing the label of the internal particles with the same sign relative to the $K^{*0}K^-K^+$ loop taking into account the minus sign of $a_1^-K^{*}\bar{K}$ and $K^*\pi K$ vertices.
Then, adding the contribution of the $K^{*-}K^0\bar{K}^0$ and $K^{*0}K^-K^+$ loops, we obtain
\begin{align}
 &-i\mM_{\Lambda_b,\Lambda_c\pi^-MM'}=-i\mM_{\Lambda_b,\Lambda_c\pi^-MM'}^{(K^{*0}K^-K^+)}-i\mM_{\Lambda_b,\Lambda_c\pi^-MM'}^{(K^{*-}K^0\bar K^0)}\notag\\
 &=-i\tilde{g}g_1G_FV_{ud}V_{cb}B_\mu G_{a_1}^{\mu\nu}\left[t_{MM',K^+K^-}L_\nu(K^{*0}K^-K^+)+t_{MM',K^0\bar K^0}L_\nu(K^{*-}K^0\bar K^0)\right].
\end{align}
In the following, we consider the case of $MM'=\pi^+\pi^-$ in the final state to see $f_0(980)$.
In the isospin symmetric case, where the isospin averaged mass and width of the mesons are used, the amplitude is reduced as follows\footnote{Note that $K^+K^-=[-(K\bar K)_{I=1}-(K\bar K)_{I=0}]/\sqrt{2}$ and $K^0\bar K^0=[(K\bar K)_{I=1}-(K\bar K)_{I=0}]/\sqrt{2}$ with a phase convention $\left|K^-\right>=-\left|I=1/2,I_z=-1/2\right>$.}:
\begin{align}
 -i\mM_{\Lambda_b,\Lambda_c\pi^-f_0}=&-2i\tilde{g}g_1G_FV_{ud}V_{cb}B_\mu G_{a_1}^{\mu\nu}L_\nu(K^{*0}K^-K^+) t_{\pi^+\pi^-,K^+K^-}.\label{eq:tmuTA1}
\end{align}
This isospin averaged amplitude will be used in the following calculation.

Using the formula of the phase space volume in Ref.~\cite{Jing:2019cbw}, the differential distribution is given by
\begin{align}
 \frac{d^2\Gamma_{\Lambda_b\rightarrow\Lambda_c\pi^-f_0}}{dM_{\pi^+\pi^-}dM_{\pi^-f_0}}=&\frac{2m_{\Lambda_c}2m_{\Lambda_b}}{2^5(2\pi)^8m_{\Lambda_b}^2}p_{\Lambda_c}p'_{\pi^-}p''_{\pi^+}\int d\Omega_{\Lambda_c}d\Omega_{\pi^-}'d{\Omega}''_{\pi^+}\overline{|\mM_{\Lambda_b,\Lambda_c\pi^-f_0}|^2},\label{eq-dddst}
\end{align}
with
\begin{align}
 p_{\Lambda_c}=&\frac{1}{2m_{\Lambda_b}}{\lambda^{1/2}(m_{\Lambda_b}^2,m_{\Lambda_c}^2,M_{\pi^-f_0}^2)},\\
 p'_{\pi^-}=&\frac{1}{{2M_{\pi^-f_0}}}{\lambda^{1/2}(M_{\pi^-f_0}^2,m_{\pi^-}^2,M_{\pi^+\pi^-}^2)},\\
 p''_{\pi^+}=&\frac{1}{{2M_{\pi^+\pi^-}}}{\lambda^{1/2}(M_{\pi^+\pi^-}^2,m_{\pi^+}^2,m_{\pi^-}^2)}.
\end{align}
The angles $\Omega_{\Lambda_c}$, $\Omega_{\pi^-}'$, and ${\Omega}_{\pi^+}''$ are those of $\Lambda_c$, $\pi^-$, and $\pi^+$ in the $\Lambda_b$ rest frame, the $\pi^-f_0(980)$ c.m. frame, and the $\pi^+\pi^-$ c.m. frame, respectively.

\section{Results\label{sec_results}}
\begin{figure}[t]
 \centering
 \includegraphics[width=15cm]{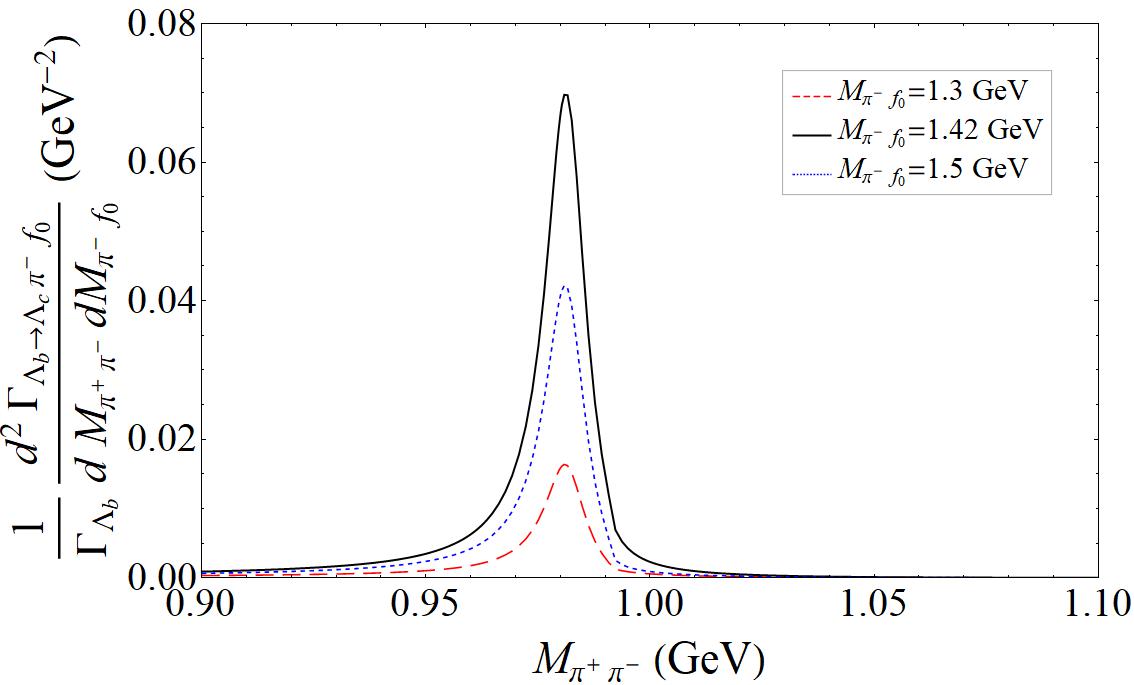}
 \caption{The $\pi^+\pi^-$ invariant mass distribution $(d^2\Gamma_{\Lambda_b\to\Lambda_c\pi^-f_0}/dM_{\pi^-f_0}dM_{\pi^+\pi^-})/\Gamma_{\Lambda_b}$.
 $M_{\pi^-f_0}$ is fixed to $1.3~\gev$ (red dashed), $1.42~\gev$ (black~solid), and $1.5~\gev$ (blue dotted).}
 \label{fig_f0-a0_1}
\end{figure}

We show in Fig.~\ref{fig_f0-a0_1} the $\pi^+\pi^-$ invariant mass distribution given by Eq.~\eqref{eq-dddst} 
normalized with the $\Lambda_b$ full width, $\Gamma_{\Lambda_b}$, 
with $M_{\pi^-f_0}=1.3$, $1.42$, and $1.5~\gev$.
A peak of $f_0(980)$ is clearly seen at $M_{\pi^+\pi^-}=0.98~\gev$, and the largest strength is given with $M_{\pi^-f_0}=1.42~\gev$ by the virtue of the $K^*\bar KK$ triangle mechanism.

\begin{figure}[t]
 \centering
 \includegraphics[width=15cm]{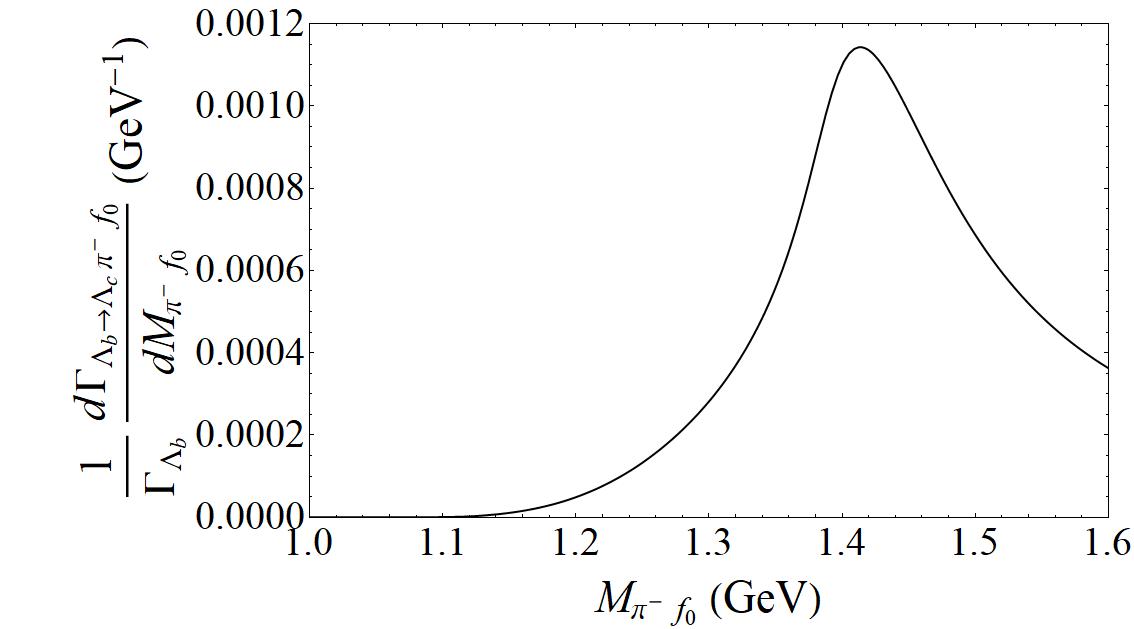}
 \caption{The $\pi^-f_0(980)$ invariant mass distribution $(d\Gamma_{\Lambda_b\to\Lambda_c\pi^-f_0}/dM_{\pi^-f_0})/\Gamma_{\Lambda_b}$ as a function of $M_{\pi^-f_0}$.}
 \label{fig_f0-a0_integrated}
\end{figure}

Integrating Eq.~\eqref{eq-dddst} over $M_{\pi^+\pi^-}$ in the range of $M_{\pi^+\pi^-}\in[0.9,1.1]~\gev$, we obtain the $\pi^-f_0(980)$ invariant mass distribution, $(d\Gamma_{\Lambda_b\to\Lambda_c\pi^-f_0}/dM_{\pi^-f_0})/\Gamma_{\Lambda_b}$, shown in Fig.~\ref{fig_f0-a0_integrated}.
The distribution is normalized with $\Gamma_{\Lambda_b}$ again.
In the $\pi^-f_0(980)$ distribution, a peak around $1.42~\gev$ with a width of the order of 0.1~GeV originating from the TS is clearly seen.
The distribution is similar to the ones in the $a_1(1260)$ decay and the $\tau$ decay calculated in Refs.~\cite{Aceti:2016yeb,Dai:2018rra}.
Note that the $\pi^+\pi^-$ pair in the final-state $\pi^-\pi^+\pi^-$ is produced by $f_0(980)$ in this calculation.
The $\pi^-\pi^+\pi^-$ in the final state would be mainly produced by the $s$-wave $\pi\rho$, which is a decay product of $a_1^-(1260)$, as studied in Refs.~\cite{Ketzer:2015tqa,Aceti:2016yeb} in the $\pi^-p\to\pi^+\pi^-\pi^-p$ reaction.

By integrating $(d\Gamma_{\Lambda_b\to\Lambda_c\pi^-f_0}/dM_{\pi^-f_0})/\Gamma_{\Lambda_b}$ in the range of $\Delta=[1.0,1.6]~\gev$, we obtain the branching ratio ${\rm Br}_\Delta$;
\begin{align}
 {\rm Br}_\Delta(\Lambda_b\rightarrow\Lambda_c\pi^-f_0(980);f_0(980)\rightarrow\pi^+\pi^-)
 \equiv&\frac{1}{\Gamma_{\Lambda_b}}\int_\Delta dM_{\pi^-f_0}\frac{d\Gamma_{\Lambda_b\to\Lambda_c\pi^-f_0}}{dM_{\pi^-f_0}}\label{eq:prdef}\\
 =&2.2\tento{-4},\label{eq:brd-f0}
\end{align}
which is the same order of magnitude obtained in the $\tau^-$ decay into $\nu_\tau\pi^-f_0(980)$ via the triangle mechanism~\cite{Dai:2018rra}.

\begin{figure}[t]
 \centering
 \includegraphics[width=15cm]{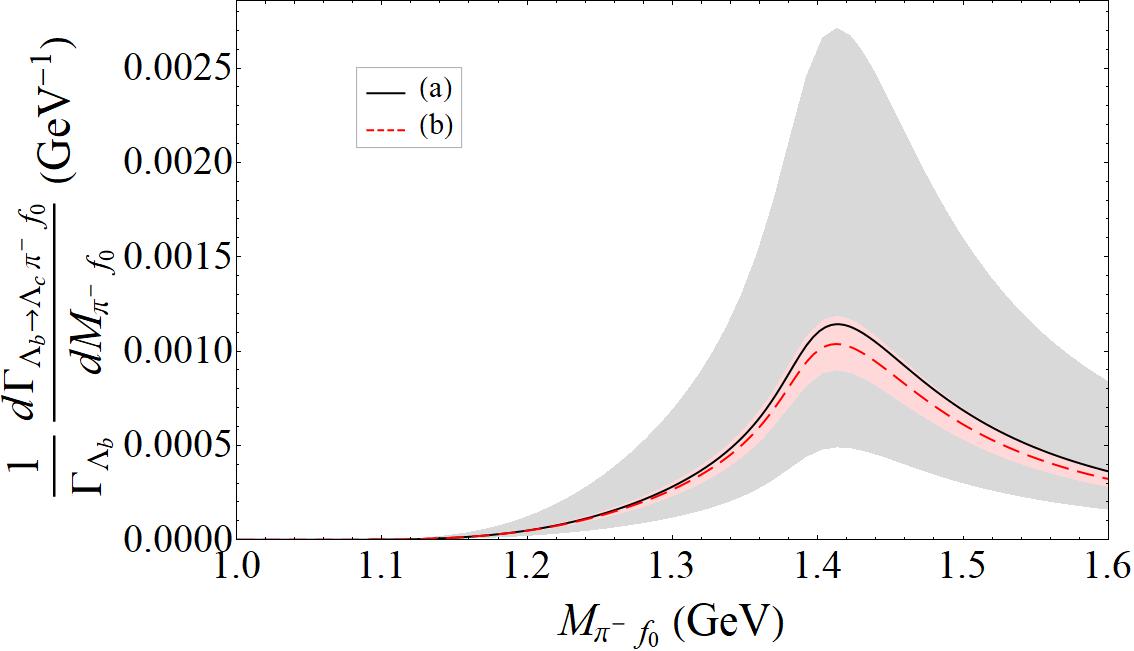}
 \caption{The $\pi^-f_0(980)$ invariant mass distribution $(d\Gamma_{\Lambda_b\to\Lambda_c\pi^-f_0}/dM_{\pi^-f_0})/\Gamma_{\Lambda_b}$ with different parameters in the $\Lambda_b\rightarrow\Lambda_c$ transition form factor. The lines (a) and (b), which are plotted with the black-solid and red-dashed curves, are the plots with the parameter set of the $\Lambda_b\to\Lambda_c$ form factor given in Ref.~\cite{Azizi:2018axf} and \cite{Faustov:2016pal}, respectively.
 The gray and red bands of the lines (a) and (b) reflect the uncertainties of the parameters in the $\Lambda_b\to\Lambda_c$ form factors in each model.}
 \label{fig8}
\end{figure}

To see the uncertainties from the $\Lambda_b\rightarrow\Lambda_c$ transition form factors, we show the plot in Fig.~\ref{fig8} with different parameter sets of the $\Lambda_b\rightarrow\Lambda_c$ form factors given in Refs.~\cite{Azizi:2018axf,Faustov:2016pal} which are denoted by the lines (a) and (b), respectively.
The gray and red bands are the uncertainties of the lines (a) and (b) originating from the parameters in the $\Lambda_b\to\Lambda_c$ form factors.
The gray band for the uncertainties of the line (a) is obtained by using the errors of $\mathcal{F}(0)$ in Eq.~\eqref{eq:ffdef} given in Ref.~\cite{Azizi:2018axf}.
In Ref.~\cite{Faustov:2016pal}, the errors of the form factors are estimated less than 5\%, 
and here the uncertainties of the line (b) expressed with the red band are given by changing $\mathcal{F}(0)$ by $\pm 5\%$.
One can see the relatively large uncertainties of the line (a) expressed with the gray band.
The branching ratio ${\rm Br}_\Delta$ is in the range from $1\tento{-4}$ to $5\tento{-4}$, still the order $10^{-4}$.
Comparing the lines (a) and (b) in Fig.~\ref{fig8}, one will see the similar line shapes with the different parameter sets; 
the peak structure around 1.42~GeV is stable.
We note that, for the $\Lambda_b\to\Lambda_c$ transition amplitude, only the external $W^-$ emission diagram is taken into account, 
and the approximation gives some further uncertainties which are not addressed in this study.

In the amplitude Eq.~\eqref{eq:lbtolckstkb}, the $a_1(1260)$ dominance in the $K^{*-}K^0$ and $K^{*0}K^-$ production is assumed.
For comparison, we show the plot without the intermediate $a_1(1260)$ resonance in the production.\footnote{In Ref.~\cite{Sakai:2017hpg}, the effects of the $a_1(1260)$ meson in the $B^-\to K^-K^{*0}D^{(*)0}$ transition part of the $B^-\to K^-\pi^-D^+_{s0(s1)}$ decay with a $K^{*0}D^{(*)0}K^+$ triangle loop are studied, and it is found that the peak originating from the triangle mechanism is not changed with the inclusion of the $a_1(1260)$ contribution.}
In terms of the $K^*\bar K$ interaction, if the $K^*\bar K$ interaction is weak or moderately attractive and the coupling to the $a_1(1260)$ is not so large, the $K^*\bar K$ rescattering amplitude is expected to have a moderate energy dependence, and the $K^*\bar K$ production from $W^-$ can be approximated with a constant contact term involving all the short-range physics of the process.
On the other hand, if the $K^*\bar K$ interaction is sufficiently strong, the $K^*\bar K$ generates a pole dynamically~\cite{Lutz:2003fm,Roca:2005nm,Zhou:2014ila}, and the $K^*\bar K$ rescattering can be represented approximately with the coupling to the pole, which may be related to $a_1(1260)$, and the details of the $K^*\bar K$ interaction is encoded in the coupling constant of the pole and the $K^*\bar K$ channel from the viewpoint of the Weinberg compositeness relation~\cite{Weinberg:1965zz}.
Comparing the $\pi^-f_0(980)$ distributions with and without $a_1(1260)$, we can see the effect of the different production mechanisms and the interaction of the initial $K^*\bar K$ pair on the $\pi^-f_0(980)$ distribution and the stability of the TS peak against it.
In the case without the intermediate $a_1(1260)$, the decay amplitude of $\Lambda_b\to\Lambda_c\pi^-f_0(980)$ is given by replacing the $a_1(1260)$ propagator $G_{a_1}^{\mu\nu}$ with $g^{\mu\nu}$ in Eq.~\eqref{eq:tmuTA1}.
The $W^-\to K^*\bar K$ amplitude is given by Eq.~\eqref{eq-v1} by replacing the $a_1^-(1260)$ polarization vector with the $W^-$ one.

\begin{figure}[t]
 \centering
 \includegraphics[width=15cm]{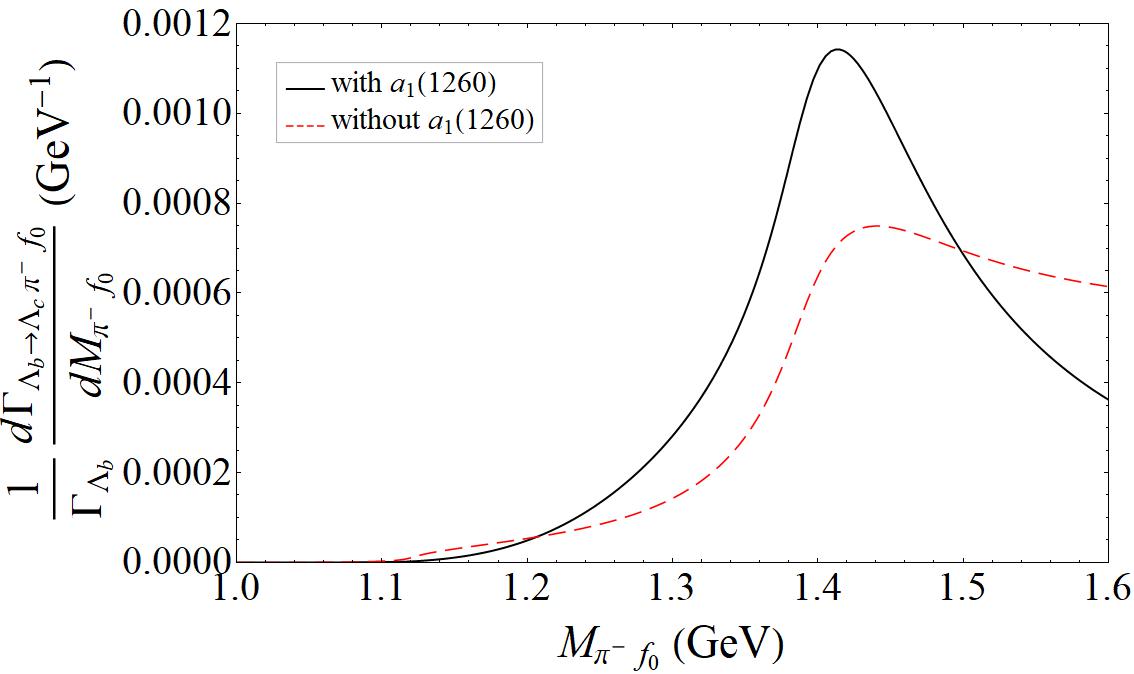}
 \caption{The $\pi^-f_0(980)$ distribution $(d\Gamma_{\Lambda_b\to\Lambda_c\pi^-f_0}/dM_{\pi^-f_0})/\Gamma_{\Lambda_b}$ with and without $a_1(1260)$.
 The black-solid (red-dashed) line is the plot with (without) the $a_1(1260)$ resonance.}
 \label{fig:noa1}
\end{figure}

The $\pi^-f_0(980)$ invariant mass distributions with and without $a_1(1260)$ are compared in Fig.~\ref{fig:noa1}.
The peak around $1.42~\gev$ can be seen in both cases,
and the distribution without $a_1(1260)$ has a longer tail than that with $a_1(1260)$.
The branching ratio defined in Eq.~\eqref{eq:prdef} is ${\rm Br}_\Delta=1.8\tento{-4}$ without the intermediate $a_1(1260)$ resonance.
The ratio is smaller compared to Eq.~\eqref{eq:brd-f0}, but it is still the same order of magnitude.

To clarify the feature of the triangle mechanism in the $\Lambda_b\to\Lambda_c\pi^-f_0(980)$ process, 
we compare the invariant mass distribution of $\pi^-f_0(980)$ produced with and without the triangle mechanism.
The amplitudes of $a_1(1260)\to\pi^-f_0(980)$ in the $p$ wave and $f_0(980)\to \pi^+\pi^-$ in the $s$ wave,
which are needed for the amplitude of the direct production of the $p$-wave $\pi^-f_0(980)$ pair from $a_1^-(1260)$, 
are written as
\begin{align}
 -it_{a_1^-,\pi^-f_0(980)}=&g_1'\epsilon_{a_1^-}\cdot p_{\pi^-},\\
 -it_{f_0(980),\pi^+\pi^-}=&ig_{f_0,\pi\pi}. 
\end{align}
The decay amplitude of $\Lambda_b\to\Lambda_c\pi^-f_0(980)$ followed by $f_0(980)\to\pi^+\pi^-$ with $\pi^-f_0(980)$ directly produced by $a_1^-(1260)$ is given by
\begin{align}
 -i\mM_{\Lambda_b,\Lambda_c\pi^-f_0}'=&-G_FV_{ud}V_{cb}B^\mu (G_{a_1})_{\mu\nu}\frac{g_{f_0,\pi\pi}g_1'p_{\pi^-}^\nu}{M_{\pi^+\pi^-}^2-m_{f_0}^2+im_{f_0}\Gamma_{f_0}}.\label{eq:pif0direct}
\end{align}
For simplicity, we just use a Breit-Wigner amplitude of the $f_0(980)$ resonance with the mass and width from the PDG~\cite{Tanabashi:2018oca}.\footnote{In the studies of the $f_0(980)$ resonance, the Flatt\'e(-like) amplitude~\cite{Flatte:1976xu} is used to analyze its properties due to the nearby $K\bar K$ threshold; see, e.g., Refs.~\cite{Baru:2003qq,Baru:2004xg,Baru:2010ww}.}

\begin{figure}[t]
 \centering
 \includegraphics[width=15cm]{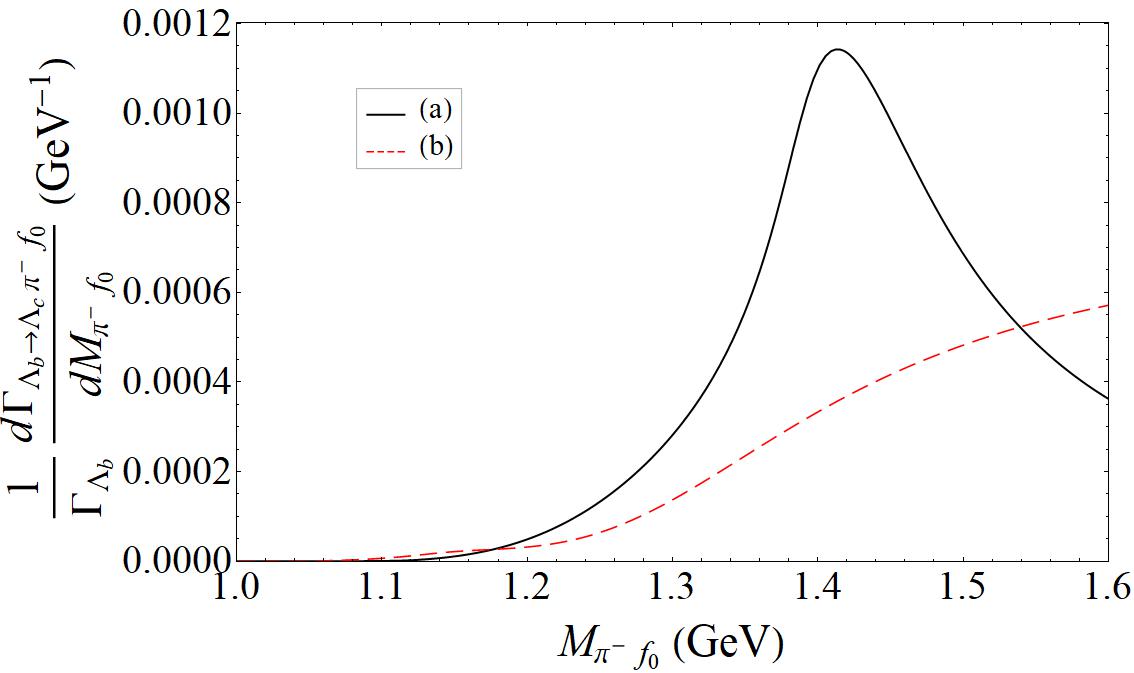}
 \caption{Comparison of the plots with different production mechanisms.
 The lines (a) and (b) are the plots with the $\pi^-f_0(980)$ pair via the triangle mechanism and the direct production by $a_1^-(1260)$, respectively.
 The amplitudes Eqs.~\eqref{eq:tmuTA1} and \eqref{eq:pif0direct} are used for the plot of lines (a) and (b), respectively.
 In the plot of the line~(b), the parameters are fixed to be the same order as the line~(a).}
 \label{fig:comp-dif-prod}
\end{figure}

In Fig.~\ref{fig:comp-dif-prod}, the $\pi^-f_0(980)$ invariant mass distribution with Eq.~\eqref{eq:pif0direct} is compared with the one with the triangle mechanism given by Eq.~\eqref{eq:tmuTA1}.
The distribution with $\pi^-f_0(980)$ directly produced by $a_1(1260)$ has no structure around 1.4~GeV and just increases as a $p$ wave.
On the other hand, in the case with the triangle loop contribution, the peak of the TS is located around 1.42~GeV with the width about 0.2~GeV in the distribution.
Thus, the triangle mechanism has the clear distinction from the other production mechanism we considered here.

\section{Conclusions\label{sec_summary}}
We have studied the $\Lambda_b\rightarrow\Lambda_c\pi^-f_0(980)$ decay with $f_0(980)\rightarrow\pi^+\pi^-$.
For the $\Lambda_b\rightarrow\Lambda_cK^*\bar{K}(K\bar{K}^*)$ production part, the amplitude is factorized into the $\Lambda_b\rightarrow\Lambda_c$ transition 
and $K^{*-}K^0(K^-K^{*0})$ production from the $a_1^-(1260)$ resonance which are connected with a $W^-$ boson taking the leading contribution in terms of the color counting~\cite{Chau:1982da}.
The $\Lambda_b\rightarrow\Lambda_c$ transition form factors are taken from the theoretical studies~\cite{Azizi:2018axf,Faustov:2016pal},
and the chiral unitary approach is employed for the $K\bar K\to \pi^+\pi^-$ transition amplitude~\cite{Liang:2014tia}.
A coupling constant related to the production of $K^{*-}K^0$ is fixed with the $\tau^-\rightarrow{\nu}_\tau K^{*-}K^0$ branching ratio assuming the $a_1(1260)$ dominance.

A peak of the $f_0(980)$ resonance is seen in the $\pi^+\pi^-$ invariant mass distribution, 
and the peak has the largest strength when the $\pi^-f_0(980)$ invariant mass is fixed to be $1.42~\gev$.
Integrating the $\pi^+\pi^-$ distribution, we obtain the $\pi^-f_0(980)$ distribution which has a peak around $1.42~\gev$ due to the triangle singularity of the $K^*\bar{K}K$ loop.
With further integration over the $\pi^-f_0(980)$ invariant mass in the range of $M_{\pi^-f_0}\in[1.0,1.6]~\gev$,
the branching ratio of $\Lambda_b\rightarrow\Lambda_c\pi^-f_0(980)$ with $f_0(980)\rightarrow\pi^+\pi^-$ by the $K^*\bar KK$ triangle mechanism is obtained as $2.2\tento{-4}$.
Considering the uncertainties from the parameters appearing in this calculation, the renormalization scale for the loop regularization, and the parameters in the $\Lambda_b\to\Lambda_c$ transition form factor, it is found that the branching ratio of $\Lambda_b\to\Lambda_c\pi^-f_0(980);f_0(980)\to\pi^+\pi^-$ is the order $10^{-4}$ and the peak position originating from the triangle mechanism is not changed
although a more sophisticated treatment of the $\Lambda_b\to\Lambda_cK^*\bar K$ transition part may be needed in the future for more definite predictions.
The comparison of the distributions with and without the intermediate $a_1(1260)$ is also done,
and it is found that the peak around 1.42~GeV is not changed even if the $a_1(1260)$ is omitted, while some difference in the shape of the distribution can be seen.
The branching ratio without the $a_1(1260)$ is still the order $10^{-4}$. 
The distribution of $\pi^-f_0(980)$ directly produced by the $a_1(1260)$ meson without the triangle loop is also considered to compare it with the distribution including the triangle loop contribution, and it is found that the distribution without the triangle loop just increases without peak structures,
which is quite different from the distribution with the $K^*\bar KK$ triangle mechanism.

The part of the $K^*\bar{K}K$ triangle loop is identical to the mechanism considered in Refs.~\cite{Ketzer:2015tqa,Aceti:2016yeb} to explain the $a_1(1420)$ peak in $\pi^-p\to \pi^-\pi^-\pi^+p$ observed by the COMPASS Collaboration~\cite{Adolph:2015pws}.
Then, future measurements of the branching ratio of the $\Lambda_b\to\Lambda_c\pi^-f_0(980)$ and the $\pi^-f_0(980)$ invariant mass distribution, particularly the peak structure around 1.4~GeV, which are the predictions made in this work, can provide a support of the $a_1(1420)$ peak as a manifestation of the triangle singularity,
and they also provide further knowledge about the role of the triangle singularities in the hadronic reactions.

\begin{acknowledgements}
 We thank Eulogio~Oset for his comments.
 S.~S. is supported in part by the National Natural Science Foundation of China (NSFC) and  the Deutsche Forschungsgemeinschaft (DFG) through the funds provided to the Sino-German Collaborative Research Center  CRC110 ``Symmetries and the Emergence of Structure in QCD''  (NSFC Grant No.~11621131001), by the NSFC under Grants No.~11835015, No.~11947302, and No.~11961141012, by the Chinese Academy of Sciences (CAS) under Grants No. QYZDB-SSW-SYS013 and No.~XDPB09, by the CAS Center for Excellence in Particle Physics (CCEPP), by the 2019 International Postdoctoral Exchange Program, and by the CAS President's International Fellowship Initiative (PIFI) under Grant No.~2019PM0108.
\end{acknowledgements}

\bibliography{biblio}
 
\end{document}